\documentclass[a4paper, 11pt, oneside]{article}
\pdfoutput=1
\usepackage{jheppub}

\usepackage{amssymb}
\usepackage{epsfig}
\usepackage{graphicx}% Include figure files
\usepackage{subfigure}
\usepackage{dcolumn}% Align table columns on decimal point
\usepackage{bm}% bold math
\usepackage[usenames ,dvipsnames]{xcolor}
\usepackage{slashed}

\newcommand{\nc}{\newcommand}
\nc{\beq}{\begin{equation}}  \nc{\eeq}{\end{equation}}
\nc{\bea}{\begin{eqnarray}}  \nc{\eea}{\end{eqnarray}}
\nc{\baa}{\begin{array}}     \nc{\eaa}{\end{array}}
\nc{\bit}{\begin{itemize}}   \nc{\eit}{\end{itemize}}
\nc{\ben}{\begin{enumerate}} \nc{\een}{\end{enumerate}}
\nc{\bce}{\begin{center}}    \nc{\ece}{\end{center}}
\nc{\bpm}{\begin{pmatrix}}   \nc{\epm}{\end{pmatrix}}
\nc{\bvt}{\begin{verbatim}}  \nc{\evt}{\end{verbatim}}

\begin{document}

%\title{Leading-Order One-Loop Contribution to Dark-Matter-Nucleon Interaction in a Scalar Dark Matter Model with Tree-Level Cancellation Mechanism}
\title{One-loop contribution to dark matter-nucleon scattering in the pseudoscalar dark matter model}

\author[2]{Duarte Azevedo,}
\author[1]{Mateusz Duch,}
\author[1]{Bohdan Grzadkowski,} 
\author[1]{Da Huang,} 
\author[1]{Michal Iglicki,} 
\author[2]{and Rui Santos}
\affiliation[1]{Faculty of Physics, University of Warsaw,
Pasteura 5, 02-093 Warsaw, Poland}
\affiliation[2]{
Centro de F\'{\i}sica Te\'{o}rica e Computacional, Faculdade de Ci\^{e}ncias, Universidade de Lisboa,
Campo Grande, Edif\'{\i}cio C8 1749-016 Lisboa, Portugal}
\emailAdd{dazevedo@alunos.fc.ul.pt}
\emailAdd{mateusz.duch@fuw.edu.pl}
\emailAdd{bohdan.grzadkowski@fuw.edu.pl}
\emailAdd{da.huang@fuw.edu.pl}
\emailAdd{michal.iglicki@fuw.edu.pl}
\emailAdd{rasantos@fc.ul.pt}

\date{\today}
\abstract{
Recent dark matter (DM) direct searches place very stringent constraints on the possible DM candidates proposed in extensions of the Standard Model.
There are however models where these constraints are avoided. One of the simplest and most striking examples comes from a straightforward  
Higgs portal pseudoscalar DM model featured with a softly broken $U(1)$ symmetry. In this model the tree-level DM-nucleon scattering cross section vanishes in the limit of zero momentum-transfer. It has also been argued that the leading-order DM-nucleon cross section appears at the one-loop level. %, which is too small to be constrained experimentally. 
In this work we have calculated the exact cross section in the zero momentum-transfer at the leading-order i.e., at the one-loop level of perturbative expansion. 
We have concluded that, in agreement with expectations, the amplitude for the scattering process is UV finite and approaches zero in the limit of vanishing DM masses. 
Moreover, we made clear that the finite DM velocity correction at tree-level is subdominant with respect to the one-loop contribution. 
Based on the analytic formulae, our numerical studies show that, for a typical choice of model parameters, the DM nuclear recoiling cross section is well 
below ${\cal O}(10^{-50}~{\rm cm}^2)$, which indicates that the DM direct detection signal in this model naturally avoids the present strong experimental limits on the cross section. 
}

%\pacs{95.35.+d, 13.85.Tp, 14.80.-j, 98.70.Sa,}
%\keywords{Dark Matter, Cosmic Rays, AMS-02 Experiment}
\maketitle

%%%%%%%%%%%%%%%%%%%%%%%%%%%%%%%%%%%%%%%%%%%%%%%%%%%%%%%%%%%%%%%%%%%%%%%%%%%%%%%%%%%%%%%%%
\section{Introduction}
\label{s1}
The nature of dark matter (DM)~\cite{PDG,Bergstrom:2012fi} is still a great mystery of modern physics. Over the last several decades, there have been many experiments to search for DM particles, 
which focus either on its direct or indirect detection~\cite{Bertone:2004pz,Feng:2010gw}. In particular, the recent XENON1T experiment~\cite{Aprile:2018dbl} placed the most stringent upper bound on 
the DM-nucleon scattering cross section. The limit constitutes a great challenge while constructing DM models.

Some of the simplest realisations of DM are SM extensions with an extra complex scalar field \cite{McDonald:1993ex, Barger:2008jx}
charged under an extra global $U(1)$ and for which the real parte of the singlet acquires a vacuum expectation value~(vev)
\cite{Barger:2008jx, Barger:2010yn, Gonderinger:2012rd, Coimbra:2013qq, Gross:2017dan, Cheng:2018ajh, Azevedo:2018oxv}. Then, the pseudo-Goldstone boson $A$ (imaginary component of the complex
scalar field)  becomes the massive DM candidate if $U(1)$ is softly broken. In turn, the DM-nucleon ($AN$) scattering, mediated by the remaining scalars of the theory, the SM-like Higgs and the extra scalar, can be naturally suppressed if the linear breaking term is removed by a $Z_2$ symmetry \cite{Barger:2010yn, Gross:2017dan}. Remarkably, the tree-level DM nuclear recoiling cross section is found to vanish in the limit of zero momentum transfer. Recently, it has been argued in Ref.~\cite{Gross:2017dan} that the leading-order contribution to DM nuclear scattering arises at the one-loop order. Based on the analysis of the asymptotic behaviour at large and small DM mass, a simple approximate formula for the one-loop DM-nucleon cross sections $\sigma_{AN}$ was suggested, which implies that the natural value of $\sigma_{AN}$ should be much smaller than the current experimental upper limits. This result shows that the DM direct detection experiments do not constrain the model, which is explicitly demonstrated in Ref.~\cite{Azevedo:2018oxv} by scanning the whole parameter space.

However, there are still several potential problems plaguing the above conclusion. Firstly, both the final results given in Ref.~\cite{Gross:2017dan} and the scans shown in Ref.~\cite{Azevedo:2018oxv} depend crucially on the aforementioned approximation. One natural question to ask is how accurate this simple approximation is. Secondly, beyond the zero momentum transfer limit, the DM-nucleon cross section is nonzero with the correction coming from the finite momentum transfer. What is the typical order of this finite momentum-transfer correction? Which contribution is dominant, the tree-level cross section or the one-loop one? 
Furthermore, it is expected that increasingly smaller scattering cross sections will be probed by the next generation of planned experiments XENONnT~\cite{Aprile:2015uzo},
LZ~\cite{Mount:2017qzi} and DARWIN~\cite{Aalbers:2016jon} (see also ~\cite{Baudis:2012ig}), which imply a demand for increasing precision in the theoretical calculations.
Therefore, in order to answer the above questions, and to be prepared for the future experimental results on direct detection,
we need to explicitly calculate both contributions analytically and numerically, which is the main motivation to the present work. A similar
strategy, although in a context of a different model, has been considered very recently in Ref.~\cite{Han:2018gej}.

The paper is organised as follows. In Sec.~\ref{Sec_model}, we briefly introduce the model and clarify our notations and conventions. Sec.~\ref{Sec_Tree} is devoted to the calculations of finite tree-level contributions coming from the finite DM velocity or finite momentum transfer. In Sec.~\ref{Sec_1L}, we show the analytic expression of the one-loop DM-nucleon cross section in the limit of zero momentum transfer. Then we show our numerical studies in Sec.~\ref{Sec_Numerical}. Finally, the short summary is given in Sec.~\ref{Sec_Conc}.

%%%%%%%%%%%%%%%%%%%%%%%%%%%%%%%%%%%%%%%%%%%%%%%%%%%%%%%%%%%%%%%%%%%%%%%%%%%%%%%%%%
\section{The model}\label{Sec_model}
We begin our discussion by specifying the Higgs-portal complex scalar DM model~\cite{Gross:2017dan,Azevedo:2018oxv}. The SM is extended by an extra complex scalar singlet $S$ which possesses an intrinsic global $U(1)$ symmetry $S \to e^{i\alpha} S$. Then we softly break this dark $U(1)$ symmetry to the residual $Z_2$ symmetry $S \to -S$ via a mass term $\mu^2 S^2 + {\rm H.c.}$. Thus, the scalar potential is given by
\begin{eqnarray}\label{potential}
{\cal V} &=& -\mu_H^2 |H|^2 -\mu_S^2 |S|^2 + \lambda_H |H|^4 + \lambda_S |S|^4 + \kappa |H|^2 |S|^2 \nonumber\\
  && + \left(\mu^2 S^2 + {\rm H.c.}\right)\,,
\end{eqnarray}
where $H$ denotes the SM Higgs doublet. Note that we can make $\mu^{2}$ real by rotating the phase of $S$. As a result, an additional dark $CP$ symmetry $S \to S^*$ of the potential (\ref{potential}) emerges. Thus, the total symmetry of this model is $Z_2 \times CP$. Note also that the SM and the scalar $S$ are coupled to each other only via the quartic scalar coupling $\kappa |H|^2 |S|^2$, which is the prominent feature of Higgs portal models. 
We will consider the case in which the scalars $H$ and $S$ have non-zero vacuum expectation values (VEVs), $\langle H \rangle = (0, v_H/\sqrt{2})^T$ and $\langle S \rangle = v_S/\sqrt{2}$. By minimizing the scalar potential in Eq.~(\ref{potential}), we obtain the following two conditions
\begin{eqnarray}\label{tadpoleTree}
-\mu_H^2 + \lambda_H v_H^2 + \frac{1}{2}\kappa v_S^2 = 0\,,\nonumber\\
-(\mu_S^2 - 2\mu^{2}) + \lambda_S v_S^2 +\frac{1}{2}\kappa v_H^2 = 0\,,
\end{eqnarray}
which can determine both field VEVs as
\begin{eqnarray}
v_H^2 = \frac{(\kappa/2)\mu^2_H - \lambda_H (\mu_S^2 - 2\mu^{2})}{(\kappa^2/4)-\lambda_H\lambda_S}\,,\quad v_S^2 = \frac{(\kappa/2)(\mu^2_S - 2\mu^{2}) - \lambda_S v_H^2}{(\kappa^2/4)-\lambda_H\lambda_S}\,.
\end{eqnarray}
These are the conditions for the EW gauge and $Z_2$ symmetries to be broken spontaneously. We will be working in the unitary gauge where the scalars fields 
are written as
\begin{eqnarray}\label{DefH}
H = \left( \begin{array}{c}
0 \\
(v_H+h)/\sqrt{2} \\
\end{array} \right)\,,\quad S = \frac{v_S+s+iA}{\sqrt{2}}\,.
\end{eqnarray}
Since the $U(1)$ symmetry is softly broken, the would-be Goldstone boson $A$ becomes massive, with a mass given by $m_A = -4\mu^{2}$. On the other hand, this particle is odd under the preserved dark $CP$ symmetry, which guarantees its stability so that it can be the DM candidate. The other two $CP$-even scalar components $h$ and $s$, with their mass squared matrix given by
\begin{eqnarray}
{\cal M}^2 = \left(\begin{array}{cc}
2\lambda_H v_H^2 & \kappa v_H v_S \\
\kappa v_H v_S & 2 \lambda_S v_S^2 \\
\end{array}\right)\,,
\end{eqnarray}
mix via the following orthogonal transformation to form the mass eigenstates $h_{1,2}$ as
\begin{eqnarray}\label{DefMH}
\left(\begin{array}{c}
h \\
s \\
\end{array}\right) = \left(\begin{array}{cc}
c_\alpha & -s_\alpha \\
s_\alpha & c_\alpha \\
\end{array}\right) \left(\begin{array}{c}
h_1 \\
h_2 \\
\end{array}\right) \,,
\end{eqnarray}
where $s_\alpha \equiv \sin \alpha$ and $c_\alpha \equiv \cos\alpha$ and $\alpha$ is the mixing angle. Denoting the masses of $h_{1,2}$ as $m_{{1,2}}$, we have the following relations
\begin{eqnarray}\label{ConsistCond}
\lambda_H = \frac{c_\alpha^2 m_1^2 + s_\alpha^2 m_2^2 }{2 v^2_H}\,,\quad \lambda_S = \frac{s_\alpha^2 m_1^2 + c_\alpha^2 m_2^2}{2 v^2_S} \,,\quad \kappa = \frac{s_\alpha c_\alpha (m_1^2 -m_2^2)}{v_H v_S}\,.
\end{eqnarray}
In the following, we will assume that $h_1$ is the SM-like Higgs already discovered already at the LHC, with a mass of 125 GeV, while $h_2$ is the other CP-even scalar which
we already know is mainly singlet-like since the bound on the mixing angle is at present of the order $|\sin \alpha| \leq 0.35$ and it come from the combined signal strength measurements of the production
and decay of the SM-like Higgs, $h_1$~\cite{Khachatryan:2016vau}. We choose as input parameters for our study $m_{1,2}$, $s_\alpha$, $v_{H,S}$ and $m_A$. The triple- and quartic-scalar terms in the scalar potential Eq.~(\ref{potential}) generate interaction vertices, which are listed in Appendix~\ref{appA}.

%%%%%%%%%%%%%%%%%%%%%%%%%%%%%%%%%%%%%%%%%%%%%%%%%%%%%%%%%%%%%%%%%%%%%%%%%%%%%%%%%%%%%%%%%%%%%%%%%%%%%%%%%%%%%%%%%
\section{Tree-level contributions to the DM-nucleon scattering at finite DM velocity} \label{Sec_Tree}
\begin{figure}[ht]
\centering
\includegraphics[scale=0.5]{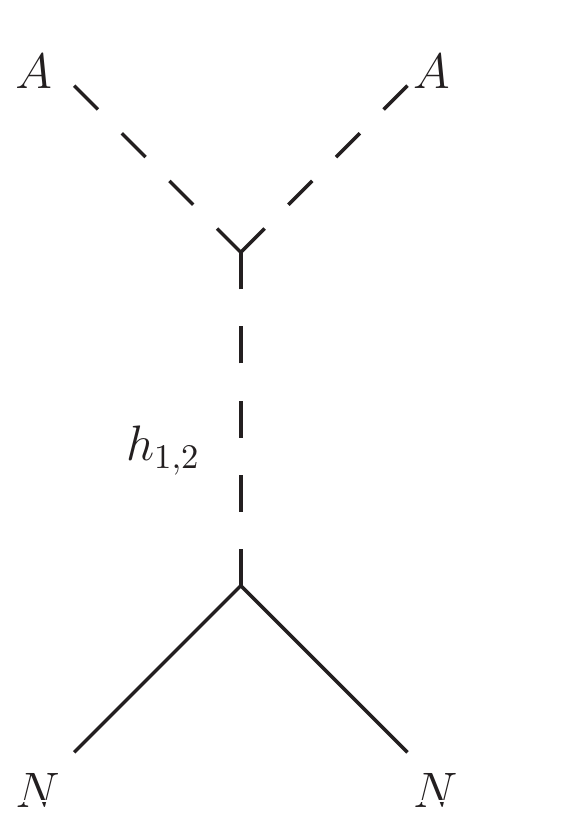}
\caption{Tree-level diagrams for DM-nucleon scattering.}\label{FigTree}
\end{figure}

At the tree-level, there are only two diagrams with $h_{1,2}$ exchange displayed in Fig.~\ref{FigTree}, the corresponding amplitude for the DM-nucleon scattering  was shown in Refs.~\cite{Gross:2017dan,Azevedo:2018oxv} to vanish in the limit of zero momentum transfer.
However, a non-vanishing DM-nucleon cross section can be obtained when we consider the finite DM velocity in the rest frame of the DM detector, although its size will be shown to be much smaller than the one-loop quantum contribution presented in the following sections.

The total amplitude for the DM-nucleon interaction at tree-level is given by
\begin{eqnarray}\label{ampTree}
-i{\cal M}_{\rm tree} &=& -\frac{i2f_N m_N}{v_H} \left(\frac{V_{AA1}c_\alpha}{q^2-m_1^2} - \frac{V_{AA2} s_\alpha}{q^2 - m_2^2}\right)\bar{u}_N(p_4) u_N(p_2)\nonumber\\
&=& -i \frac{s_\alpha c_\alpha f_N m_N}{v_H v_S} \left(\frac{m_1^2}{q^2 - m_1^2}- \frac{m_2^2}{q^2 -m_2^2}\right) \bar{u}_N(p_4) u_N(p_2)\nonumber\\
&\approx & -i \frac{s_\alpha c_\alpha f_N m_N}{v_H v_S} \left(\frac{m_1^2 -m_2^2}{m_1^2 m_2^2}\right) q^2 \bar{u}_N(p_4) u_N(p_2)\,,
\end{eqnarray}
where $q^2$ is the DM momentum transfer when it scatters against nucleons. Here $m_N$ and $f_N \approx 0.3$ represent the nucleon mass and its SM Higgs coupling~\cite{Cline:2013gha,Alarcon:2011zs,Ling:2017jyz}, respectively. In the third line, we have only kept the leading-order dependence on the momentum transfer. %due to its suppression from the relative velocity in the $AN$ system. 
Therefore, the tree-level cross section $\sigma^{\rm tree}_{AN}$ is given by
\begin{eqnarray}\label{XStree}
\sigma^{\rm tree}_{AN} \approx \frac{4 s_\alpha^2 c_\alpha^2 f_N^2}{3\pi} \frac{m_N^2 \mu_{AN}^6}{m_A^2 v_H^2 v_S^2} \frac{(m_1^2 -m_2^2)^2}{m_1^4 m_2^4} v_A^4\,,
\end{eqnarray}
where $\mu_{AN} \equiv m_A m_N/(m_A + m_N)$ is the reduced mass in the DM-nucleon system, and $v_A$ is the DM velocity in the lab frame. Note that the typical relative velocity of a DM particle in the vicinity of the Earth is expected to be $v_A \sim 200$~km/s, which would suppress the DM nuclear recoil cross section by a factor of order of $v_A^4 \sim 10^{-13}$. If we adopt typical values of parameters, say  $v_S = 1~{\rm TeV}$, $m_2 = 300~{\rm GeV}$, $s_\alpha = 0.1$ and $m_A = 100$~GeV, then the tree-level cross section is estimated as $\sigma^{\rm tree}_{AN} = 7.6\sim 10^{-68}~{\rm cm^2}$, which is too small to be observed experimentally at present but also
in the future planned experiments.

%%%%%%%%%%%%%%%%%%%%%%%%%%%%%%%%%%%%%%%%%%%%%%%%%%%%%%%%%%%%%%%%%%%%%%%%%%%%%%%%%%%%%%%%%%%%%%%%%%%%%%%%%%%%%%%%%
\section{Explicit calculation of the one-loop DM-nucleon amplitude at zero-momentum transfer}\label{Sec_1L}
\begin{figure}[ht]
\centering
\includegraphics[scale=0.5]{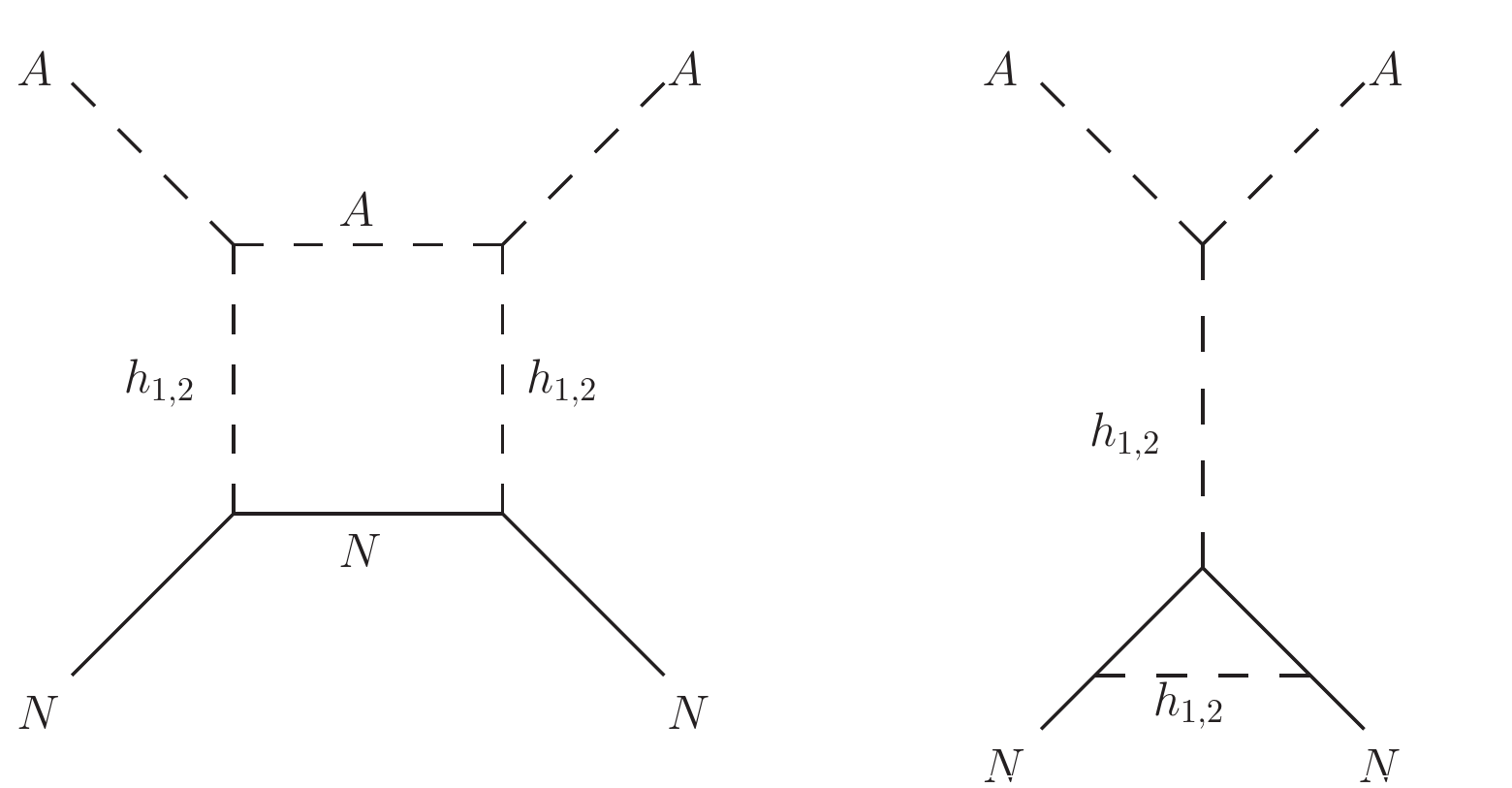}
\caption{Examples of one-loop box and nucleon-$h_{1,2}$ vertex corrected diagrams for DM-nucleon scattering, which are discarded due to the presence of the nuclear factor 
$f_N m_N/v_H \sim 1.2\times 10^{-3}$ in the nucleon-Higgs coupling.}\label{FigLN}
\end{figure}

In this section, we explicitly calculate the one-loop contributions to the DM-nucleon scattering cross section. Let us begin by noting that the nucleon couplings to both Higgs bosons $h_{1,2}$ in the model are greatly suppressed by the nuclear factor $f_N m_N/v_H \sim 1.2\times 10^{-3}$. Therefore, the leading-order contributions are given by the Feynman diagrams with only one insertion of this factor, while those with multiple insertions, such as the diagrams with box topology or with one-loop nucleon-$h_{1,2}$ vertex corrections as illustrated in Fig.~\ref{FigLN} can be discarded as they will be about three orders of magnitude smaller.
 On the other hand, it is easy to see that the
corrections with one nucleon factor insertion for the external nucleon
state and the nucleon-nucleon-Higgs vertices are always proportional
to the tree-level diagrams, so that they are cancelled identically. As
a result, the other diagrams with one insertion of nuclear factor can
be viewed as the one-loop correction $V^{(1)}_{AA1,\,AA2}$  to the
vertices $AAh_1$ and $AAh_2$.
%Note that the diagrams with one insertion of nuclear factor can be viewed as the one-loop corrections $V^{(1)}_{AA1,\, AA2}$ to the vertices $AAh_1$ and $AAh_2$. 
Furthermore, we will work in the limit of zero momentum transfer $q^2 \to 0$ in order to simplify our calculation, which is justified by the fact that the terms proportional to $q^2$ are suppressed further by powers of the relative DM velocities as previously was illustrated in the case of the tree-level computations. As a result, the one-loop contributions to the DM nuclear recoil reactions in the present model % at the leading-order of nucleon-Higgs coupling $f_N m_N/v_H$ 
can be represented as
\begin{eqnarray}\label{XS1L}
\sigma^{(1)}_{AN} = \frac{f_N^2}{\pi v_H^2} \frac{m_N^2 \mu_{AN}^2}{m_A^2} {\cal F}^2\,,
\end{eqnarray}
where the one-loop function ${\cal F}$ is defined as
\begin{eqnarray}
{\cal F} = \frac{V_{AA1}^{(1)} c_\alpha}{m_1^2} - \frac{V_{AA2}^{(1)} s_\alpha}{m_2^2}
\end{eqnarray}
with $V_{AA1\,,AA2}^{(1)}$ as the aforementioned one-loop corrections to the vertices $h_1 A^2$ and $h_2 A^2$. Therefore, our main task in the present section is to calculate the function ${\cal F}$ and associated $V^{(1)}_{AA1\,,AA2}$.

The above one-loop function, ${\cal F}$, should satisfy two consistency conditions. Firstly, since the tree-level $AN$ recoiling amplitude vanishes in the limit of zero momentum transfer, the one-loop amplitude and ${\cal F}$ should be finite in the same limit. In other words, we do not need to renormalise the model, that is, although we will define a set of counterterms, it will be shown that no renormalisation prescription is needed because the set of diagrams with counterterms only is zero. Consequently, the sum of all diagrams has to be finite.
Secondly, in the limit of $m_A^2 = -4\mu^2
\to 0$, the dark matter particle $A$ would return to its true Goldstone boson nature due to the spontaneous breaking of the global $U(1)$ symmetry. In this limit, it is argued in Ref.~\cite{Gross:2017dan} that the corresponding $AN$ scattering amplitude should be only proportional to $q^2$ and thus it should vanish when $q^2 \to 0$. This indicates that ${\cal F}$ should approach zero in the limit $m_A^2 \to 0$. These observations are two important criteria, which are useful to check the correctness of our final result.

%%%%%%%%%%%%%%%%%%%%%%%%%%%%%%%%%%%%%%%%%%%%%%%%%%%%%%%%%%%%%%%%%%%%%%%%%%%%%%%%%%%%%%%%%%%%%%%%%%%%%%%%%%%%%%%%%
\subsection{Counterterms and the cancellation of the counterterm-insertion diagrams}
\begin{figure}[ht]
\centering
\includegraphics[width=\textwidth]{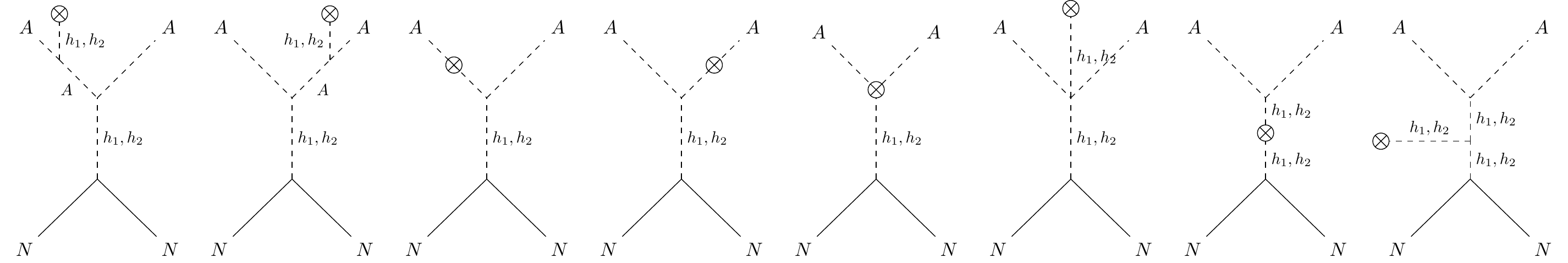}
\caption{Counterterm-insertion diagrams.}\label{FigLC}
\end{figure}

Before delving into the calculation of the one-loop diagrams, we firstly show the cancellation of the counterterm-insertion diagrams in Fig.~\ref{FigLC}. In order to do that, we need to specify some relations among the counterterms in the present DM model.

The model has 6 independent parameters as can be seen from Eq.~(\ref{potential}) that defines the potential and therefore we need 6 counterterms to cancel the UV divergences at the one-loop order. Note that since we will show that we do not need a renormalisation prescription, we refrain from discussing the complete set of renormalisation constants in the Lagrangian. The contribution from the remaining SM terms will be discussed later. We have two methods to construct such counterterms. One is to work with the original Lagrangian parameters in Eq.~(\ref{potential}) with the following counterterms %. The counterterms can be obtained by just splitting the original scalar potential as follows
\beq\label{ct1}
{\cal V}_c = -\delta\mu_H^2 |H|^2 -\delta\mu_S^2 |S|^2 + \delta\lambda_H |H|^4 + \delta\lambda_S |S|^4 + \delta\kappa |H|^2 |S|^2
   + \left({\delta\mu^{2}} S^2 + {\rm H.c.}\right)\,,
\eeq
which takes the same form as the original potential with the subscript $c$ labelling counterterms. Here we have assumed that the parameters in Eq.~(\ref{potential}) correspond to the renormalized quantities. Furthermore, as we shall see below, we do not need the field wave function renormalisation counterterms since their contributions either vanish in the limit of zero momentum transfer or are cancelled in the computations\footnote{In fact, we could have opted to work with unrenormalised fields~\cite{Sirlin:1980nh} which gives rise to Green functions that are in general divergent but leads to finite S-matrix elements.}.
The other way is to define them in terms of the physical mass eigenstates $h_{1,\,2}$ and $A$, by writing the following potential terms up to quadratic ones
\beq\label{ct2}
{\cal V}_c^{(2)} = \delta t_1 h_1 + \delta t_2 h_2 + \frac{1}{2} \delta m_1^2 h_1^2 + \frac{1}{2} \delta m_2^2 h_2^2 + \delta m^2_{12} h_1 h_2 + \frac{1}{2} \delta m_A^2 { A^2} \,.
\eeq
%\bg{something is missing? do we need $\delta m_{12}^2$? I guess $A^2$ is missing?}
These two sets of counterterms can be related to each other by expanding Eq.~(\ref{ct1}) in terms of the mass eigenstates, Eq.~(\ref{DefMH}), up to quadratic order in fields. The original set can be written
in terms of the new set of parameters as
\begin{eqnarray}\label{ctRe}
\delta \mu_H^2 &=& \frac{1}{2} (c_\alpha^2 \delta m_1^2 + s_\alpha^2 \delta m_2^2 - 2 s_\alpha c_\alpha \delta m_{12}^2) + \frac{v_S}{2v_H}[s_\alpha c_\alpha (\delta m_1^2-\delta m_2^2)+(c_\alpha^2-s_\alpha^2)\delta m_{12}^2]\nonumber\\
&& - \frac{3}{2v_H}(\delta t_1 c_\alpha - \delta t_2 s_\alpha)\,,\nonumber\\
\delta \mu_S^2 &=& \frac{1}{2} (s_\alpha^2 \delta m_1^2 + c_\alpha^2 \delta m_2^2 + 2 s_\alpha c_\alpha \delta m_{12}^2-\delta m_A^2) + \frac{v_H}{2v_S}[s_\alpha c_\alpha (\delta m_1^2-\delta m_2^2)+(c_\alpha^2-s_\alpha^2)\delta m_{12}^2]\nonumber\\
&& - \frac{1}{v_S}(\delta t_1 s_\alpha + \delta t_2 c_\alpha)\,,\nonumber\\
\delta \mu^{2} &=& \frac{1}{4v_S} (\delta t_1 s_\alpha + \delta t_2 c_\alpha) - \frac{1}{4} \delta m_A^2 \,,\nonumber\\
\delta \kappa &=&  \frac{1}{v_H v_S} [s_\alpha c_\alpha (\delta m_1^2 - \delta m_2^2) + (c^2_\alpha - s_\alpha^2)\delta m_{12}^2]\,,\nonumber\\
\delta \lambda_H &=& \frac{1}{2v_H^2} (c_\alpha^2 \delta m_1^2 + s_\alpha^2 \delta m_2^2 -2 s_\alpha c_\alpha \delta m_{12}^2) - \frac{1}{2v_H^3} (\delta t_1 c_\alpha -\delta t_2 s_\alpha)\,,\nonumber\\
 \delta \lambda_S &=& \frac{1}{2v_S^2} (s_\alpha^2 \delta m_1^2 + c_\alpha^2 \delta m_2^2 + 2 s_\alpha c_\alpha \delta m_{12}^2) - \frac{1}{2v_S^3} (\delta t_1 s_\alpha + \delta t_2 c_\alpha)\,.
\end{eqnarray}

Now we proceed to compute the total contribution from the counterterm insertion diagrams shown in Fig.~\ref{FigLC}. Note that the diagrams with external $A$-line corrections imply the following contributions to ${\cal F}$
\begin{eqnarray}
{\cal F}_{ce} = -2\left(\delta_A p^2 -\delta m_A^2+\frac{2V_{AA1}\delta t_1}{m_1^2} + \frac{2V_{AA2}\delta t_2}{m_2^2}\right)\frac{1}{p^2 - m_A^2} {\cal F}_0 =0\,,
\end{eqnarray}
where the subscript $e$ represents the external DM lines. Here $\delta_A$ is the DM $A$ wave function counterterm and $p^2$ is its momentum, and
\begin{eqnarray}
{\cal F}_0 = \frac{V_{AA1} c_\alpha}{m_1^2} -\frac{V_{AA2} s_\alpha}{m_2^2}\,
\end{eqnarray}
is the tree-level counterpart of ${\cal F}$ which appears in the first equality of Eq.~(\ref{ampTree}) in the limit of zero momentum transfer. Note that ${\cal F}_0 = 0$ if we apply the tree-level relations in Eq.~(\ref{ConsistCond}), which leads to the vanishing ${\cal F}_{ce}$. 
%\bg{I am not sure what are you saying} 
%Obviously, this contribution is proportional to the tree-level contribution which is represented in the parenthesis in the first equality. Thus, it vanishes as the tree-level one does. 
For the remaining diagrams in  in Fig.~\ref{FigLC}, we can calculate their contributions to the effective vertices $AAh_1$ and $AAh_2$ directly as
\begin{eqnarray}\label{Vcb}
-i V_{AA1\,c(i+v)}^{(1)} &=& iV_{AA1} \frac{\delta m_1^2}{m_1^2} + iV_{AA2} \frac{\delta m_{12}^2}{m_2^2} \nonumber\\
&&  -\frac{6iV_{AA1}V_{111} \delta t_1}{m_1^4} -\frac{2 i V_{AA1} V_{112} \delta t_2}{m_1^2 m_2^2}  -\frac{2 i V_{AA2} V_{112} \delta t_1}{m_1^2 m_2^2} - \frac{2i V_{AA2} V_{122} \delta t_2}{m_2^4} \nonumber\\
&& +\frac{2i V_{AA11}\delta t_1}{m_1^2} + \frac{iV_{AA12}\delta t_2}{m_2^2} -i(s_\alpha v_S \delta \lambda_S + \frac{1}{2} c_\alpha v_H \delta \kappa)\,, \nonumber\\
-iV_{AA2\,c(i+v)}^{(1)} &=& iV_{AA1} \frac{\delta m_{12}^2}{m_1^2} + iV_{AA2} \frac{\delta m_2^2}{m_2^2}\nonumber\\
&&-\frac{2iV_{AA1} V_{112} \delta t_1}{m_1^4} - \frac{2iV_{AA1}V_{122}\delta t_2}{m_1^2 m_2^2} -\frac{2iV_{AA2}V_{122} \delta t_1}{m_1^2 m_2^2} - \frac{6iV_{AA2}V_{222}\delta t_2}{m_2^4} \nonumber\\
&&+\frac{iV_{AA12}\delta t_1}{m_1^2} + \frac{2iV_{AA22}\delta t_2}{m_2^2} -i(c_\alpha v_S \delta \lambda_S - \frac{1}{2} s_\alpha v_H \delta \kappa)\,,
\end{eqnarray}
where the subscripts $i$ and $v$ denote the corrections to internal $h_{1,2}$ propagators and $h_{1,2} A^2$ vertices, respectively. We also set the four-momenta of the internal $h_{1,2}$ lines to be zero since the momentum transfer vanishes by assumption. In each equation in Eq.~(\ref{Vcb}), the first two lines correspond to the internal $h_{1,2}$ propagator corrections, while the third line to the vertices $AAh_1$ and $AAh_2$ corrections. With these two expressions, we can show their contributions to ${\cal F}$ vanishes
\begin{eqnarray}
{\cal F}_{c(i+v)} = \frac{V^{(1)}_{AA1\, cb} c_\alpha}{m_1^2} - \frac{V^{(1)}_{AA2\, cb}s_\alpha}{m_2^2} =0\,,
\end{eqnarray}
where we have used the relations in Eq.~(\ref{ctRe}) to represent the dimensionless coupling counterterms $\delta \kappa$, $\delta \lambda_S$, and $\delta \lambda_H$ in terms of the ones defined with physical mass eigenstates. We have also employed the definitions of the tree-level vertices Eq.~(\ref{vertices}) and the relations in Eq.~(\ref{ConsistCond}).

%%%%%%%%%%%%%%%%%%%%%%%%%%%%%%%%%%%%%%%%%%%%%%%%%%%%%%%%%%%%%%%%%%%%%%%%%%%%%%%%%%%%%%%%%%%%%%%%%%%%%%%%%%%%%%%%%
\subsection{Cancellation of SM Particle Loops}
\begin{figure}[ht]
\centering
\includegraphics[width=0.8\textwidth]{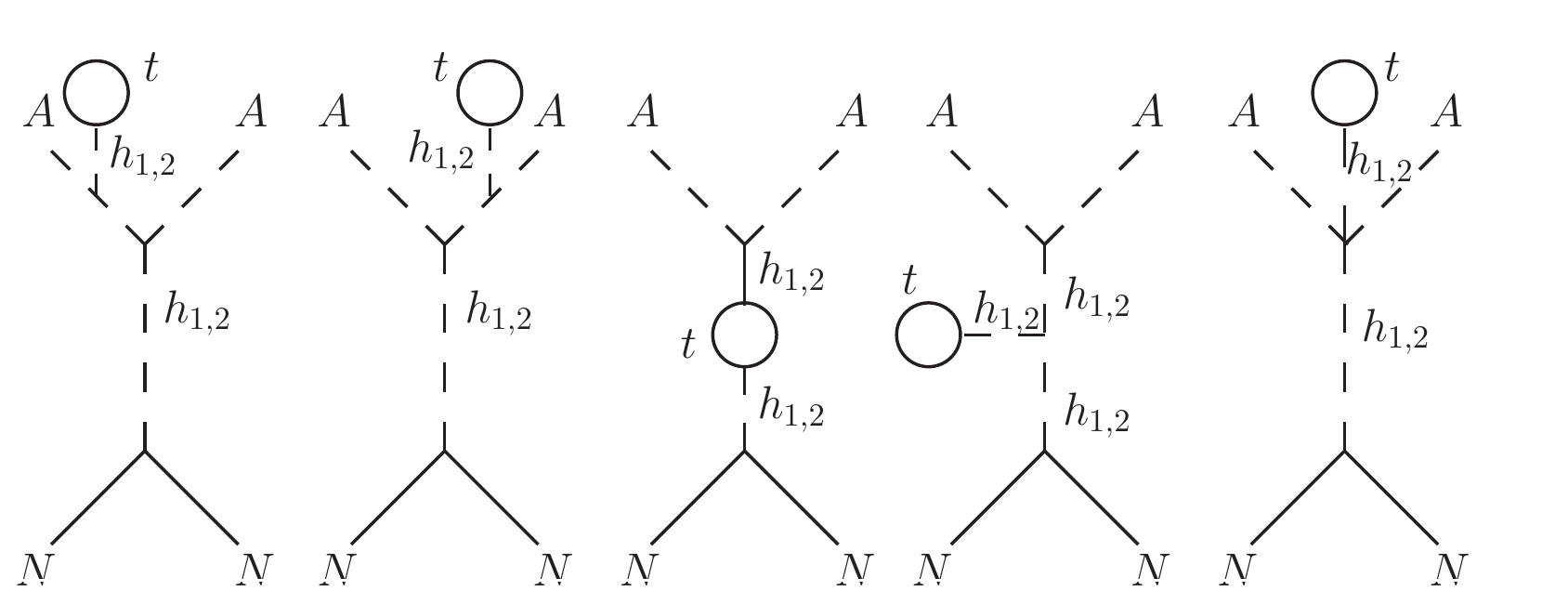}
\caption{Top quark loop diagrams for DM-nucleon scattering.}\label{FigLSM}
\end{figure}

In this subsection, we will show that the one-loop contributions from the SM particle loops other than the Higgs cancels.
For illustration purposes we will adopt the top-quark loops in Fig.~\ref{FigLSM} to show the main features of this cancellation.
Note that the remaining SM particles, quarks, leptons, and electroweak gauge bosons, couple to the Higgs bosons $h_{1,2}$ only through the rotation of the doublet neutral components $h$ with the couplings given by
\begin{eqnarray}
g_{\eta 1} = g_\eta c_\alpha \,,\quad g_{\eta 2} = -g_\eta s_\alpha\,,
\end{eqnarray}
where $g_\eta$ represents the SM particle species $\eta$ coupling to the original SM Higgs $h$. For the top quark, its couplings to $h_{1,2}$ are $y_{t1} = y_t c_\alpha$ and $y_{t2} = -y_t s_\alpha$, respectively. Moreover, it can be seen from Fig.~\ref{FigLSM} that the SM loops can appear in corrections via the Higgs bosons tadpoles, either connected to the dark matter particle A or to another Higgs line, 
or via two-point functions, which are corrections to the Higgs propagators or finally as corrections to vertices. 
%It is easy to obtain the contribution to ${\cal F}$ from the external $A$ correction as follows
%Obviously, the contribution to ${\cal F}$ from the external $A$ correction is proportional to the tree-level diagrams, indicating that it vanishes when the momentum transfer goes to zero. 
For these three contributions, the top-quark-loop $AAh_1$ and $AAh_2$ corrections are given by
\begin{eqnarray}
-i V^{(1)}_{AA1\,e} &=& -\frac{2V_{AA1}}{p^2-m_A^2} \left(\frac{V_{AA1}c_\alpha}{m_1^2} + \frac{V_{AA2}s_\alpha}{m_s^2} \right)L_1\,,\nonumber\\
-i V^{(1)}_{AA2\,e} &=& -\frac{2V_{AA2}}{p^2-m_A^2} \left(\frac{V_{AA1}c_\alpha}{m_1^2} + \frac{V_{AA2}s_\alpha}{m_s^2} \right)L_1\,,\nonumber
\end{eqnarray}
\begin{eqnarray}
-i V^{(1)}_{AA1\,i} &=& -\left(\frac{V_{AA1}c_\alpha^2}{m_1^2} - \frac{V_{AA2}c_\alpha s_\alpha}{m_2^2} \right) L_2 \nonumber\\
&& + \left( \frac{6V_{AA1}V_{111} c_\alpha}{m_1^4} - \frac{2V_{AA1}V_{112}s_\alpha}{m_1^2 m_2^2} + \frac{2V_{AA2}V_{112}c_\alpha}{m_1^2 m_2^2} - \frac{2V_{AA2}V_{122}s_\alpha}{m_2^4} \right) L_1\,,\nonumber\\
-i V^{(1)}_{AA2\,i} &=& -\left(-\frac{V_{AA1}s_\alpha c_\alpha}{m_1^2} + \frac{V_{AA2} s_\alpha^2}{m_2^2}\right) L_2 \nonumber\\
&& + \left(\frac{2V_{AA1}V_{112} c_\alpha}{m_1^4} - \frac{2V_{AA1}V_{122}s_\alpha}{m_1^2 m_2^2} + \frac{2V_{AA2}V_{122}c_\alpha}{m_1^2 m_2^2} -\frac{6V_{AA2}V_{222}s_\alpha}{m_2^4} \right) L_1\,,\nonumber\\
-i V^{(1)}_{AA1\,v} &=& -\left( \frac{2V_{AA11} c_\alpha}{m_1^2} - \frac{V_{AA12}s_\alpha}{m_2^2} \right)L_1,,\nonumber\\
-i V^{(1)}_{AA2\,v} &=& -\left( \frac{V_{AA12} c_\alpha}{m_1^2} - \frac{2V_{AA22}s_\alpha}{m_2^2} \right)L_1,,
\end{eqnarray}
where, for top quarks, the tadpole and bubble one-loop integrals can be represented as follows
\begin{eqnarray}
L_1 &=& (-1) (-iy_t) \int\frac{d^4l}{(2\pi)^4} {\rm Tr}\left[\frac{i}{\slashed{l}-m_t}\right]\,,\nonumber\\
L_2 &=& (-1) (-iy_t)^2 \int\frac{d^4 l }{(2\pi)^4} {\rm Tr}\left[\frac{i^2}{(\slashed{l}-m_t)^2}\right]\,,
\end{eqnarray}
where Tr denotes the trace over the spinor space. With $V^{(1)}_{AA1(AA2)\,e}$, it is easy to write down the following contribution to ${\cal F}$ from the external $A$ correction
\begin{eqnarray}
{\cal F}_e = (-i)\frac{2L_1}{p^2-m_A^2} \left(\frac{V_{AA1}c_\alpha}{m_1^2} + \frac{V_{AA2}s_\alpha}{m_2^2}\right) {\cal F}_0 =0\,,
\end{eqnarray}
in which the second equality follows the identity ${\cal F}_0 =0$. For the remaining diagrams, we can apply the definitions of the tree-level couplings in Appendix~\ref{appA} and the tree-level relations in Eq.~(\ref{ConsistCond}) to directly prove 
\begin{eqnarray}
{\cal F}_{i+v} = \frac{(V^{(1)}_{AA1\,i} + V^{(1)}_{AA1\,v})c_\alpha}{m_1^2} - \frac{(V^{(1)}_{AA2\,i} + V^{(1)}_{AA2\,v})s_\alpha}{m_2^2} =0\,.
\end{eqnarray}

In the above derivation, what is crucial for the cancellation is the dependence of top-quark Yukawa couplings on the mixing angle $\alpha$. Since for a given Higgs boson $h_i$ the mixing matrix enters the same way for all SM fermions and electroweak gauge bosons, therefore the cancellation is present for all SM particles (except $h_{1,2}$) in the loops as well.

%It is noticed that the above derivation does not depends on any specific top-quark-loop feature, which is only contained in the expressions of $L_1$ and $L_2$. Therefore, this cancellation can be also applied to other SM particles.

%%%%%%%%%%%%%%%%%%%%%%%%%%%%%%%%%%%%%%%%%%%%%%%%%%%%%%%%%%%%%%%%%%%%%%%%%%%%%%%%%%%%%%%%%%%%%%%%%%%%%%%%%%%%%%%%%
\subsection{One-Loop Level DM-Nucleon Scattering}
Having proved the cancellation of all diagrams involving the counterterms and the SM particle loops, we now focus on loop diagrams generated by the Higgs bosons $h_{1,2}$ and the scalar DM particle $A$. 
As shown below, we can divide these one-loop diagrams into three classes: the corrections to the external DM lines $A$, to the vertices $V_{AA1, AA2}$,  and to the internal Higgs propagators. Note
that all expression will be written as a function of the triple- and quartic-scalar terms in the scalar potential Eq.~(\ref{potential}) listed in Appendix~\ref{appA}.
It is useful to first define the following one-particle irreducible (1PI) one-loop diagrams.\\
\noindent $\bullet$ The $h_{1,2}$ and $A$ tadpole corrections:
\begin{eqnarray}
-i\Delta t_{1} &=& \int\frac{d^4l}{(2\pi)^4} \left(\frac{3V_{111}}{l^2-m_1^2} +\frac{V_{122}}{l^2-m_2^2} + \frac{V_{AA1}}{l^2-m_A^2} \right)\,,\nonumber\\
-i\Delta t_2 &=& \int \frac{d^4 l}{(2\pi)^4} \left( \frac{V_{112}}{l^2-m_1^2} + \frac{3V_{222}}{l^2-m_2^2} + \frac{V_{AA2}}{l^2-m_A^2} \right)\,,
\end{eqnarray}
\noindent $\bullet$ The $h_{1,2}$ and $A$ mass squared corrections:
\begin{eqnarray}
-i\Delta m_1^2 &=& \int\frac{d^4 l}{(2\pi)^4} \left[\frac{18V_{111}^2}{(l^2-m_1^2)^2} + \frac{4 V_{112}^2}{(l^2 -m_1^2)(l^2-m_2^2)} + \frac{2V_{122}^2}{(l^2 -m_2^2)^2} + \frac{2 V_{AA1}^2}{(l^2 -m_A^2)^2} \right]\nonumber\\
&& +\left[ \frac{12V_{1111}}{l^2-m_1^2} + \frac{2V_{1122}}{l^2 -m_2^2} + \frac{2V_{AA11}}{l^2-m_A^2} \right]\,,\nonumber\\
-i\Delta m_2^2 &=& \int\frac{d^4 l}{(2\pi)^4} \left[ \frac{2V_{112}^2}{(l^2 -m_1^2)^2} + \frac{4V_{122}^2}{(l^2-m_1^2)(l^2-m_2^2)} + \frac{18 V_{222}^2}{(l^2-m_2^2)^2} +\frac{2V_{AA2}^2}{(l^2-m_A^2)^2}  \right]\nonumber\\
&& +\left[ \frac{2V_{1122}}{l^2-m_1^2} + \frac{12 V_{2222}}{l^2 -m_2^2} + \frac{2V_{AA22}}{l^2-m_A^2} \right]\,,\nonumber\\
-i\Delta m_{12}^2 &=& \int\frac{d^4l}{(2\pi)^4} \left[ \frac{6V_{111}V_{112}}{(l^2-m_1^2)^2} + \frac{4V_{112}V_{122}}{(l^2-m_1^2)(l^2-m_2^2)} + \frac{6V_{122}V_{222}}{(l^2-m_2^2)^2} + \frac{2V_{AA1}V_{AA2}}{(l^2-m_A^2)^2} \right]\nonumber\\
&& + \left[ \frac{3V_{1112}}{l^2-m_1^2} + \frac{3V_{1222}}{l^2-m_2^2} + \frac{V_{AA12}}{l^2-m_A^2} \right]\,,\nonumber\\
-i\Delta m_A^2 &=& \int\frac{d^4l}{(2\pi)^4} \left[\frac{4V_{AA1}^2}{[(l+p)^2-m_A^2](l^2-m_1^2)} + \frac{4V_{AA2}^2}{[(l+p)^2-m_A^2](l^2-m_2^2)} \right]\nonumber\\
&& +\left[\frac{2V_{AA11}}{l^2-m_1^2} + \frac{2V_{AA22}}{l^2-m_2^2} + \frac{12V_{AAAA}}{l^2-m_A^2}\right] \,,
\end{eqnarray}
\noindent $\bullet$ The 1PI vertex corrections:
\begin{eqnarray}
-i\Delta V_{AA1} &=& \int\frac{d^4l}{(2\pi)^4}\left[ \frac{6V_{111}V_{AA11}}{(l^2-m_1^2)^2} + \frac{2V_{112}V_{AA12}}{(l^2-m_1^2)(l^2-m_2^2)} + \frac{2V_{122}V_{AA22}}{(l^2-m_2^2)^2} + \frac{12V_{AA1}V_{AAAA}}{(l^2-m_A^2)^2} \right]\nonumber\\
&& +2\times\left[\frac{4V_{AA1}V_{AA11}}{[(l+p)^2-m_A^2](l^2-m_1^2)} + \frac{2V_{AA2}V_{AA12}}{[(l+p)^2-m_A^2](l^2-m_2^2)}\right] \nonumber\\
&& +\left[\frac{12 V_{111}V_{AA1}^2}{[(l+p)^2-m_A^2](l^2-m_1^2)^2} + \frac{2\times 4 V_{112}V_{AA1} V_{AA2}}{[(l+p)^2-m_A^2](l^2-m_1^2)(l^2-m_2^2)} \right. \nonumber\\
&&\left. +  \frac{4 V_{122}V_{AA2}^2}{[(l+p)^2-m_A^2](l^2-m_2^2)^2} \right]\nonumber\\
&& + \left[\frac{4V_{AA1}^3}{[(l+p)^2-m_A^2]^2(l^2-m_1^2)} + \frac{4V_{AA1}V_{AA2}^2}{[(l+p)^2-m_A^2]^2 (l^2-m_2^2)}\right]\,,\nonumber
\end{eqnarray}
\begin{eqnarray}
-i\Delta V_{AA2} &=& \int\frac{d^4l}{(2\pi)^4}\left[ \frac{2V_{112}V_{AA11}}{(l^2-m_1^2)^2} + \frac{2V_{122}V_{AA12}}{(l^2-m_1^2)(l^2-m_2^2)} + \frac{6V_{222}V_{AA22}}{(l^2-m_2^2)^2} + \frac{12V_{AA2}V_{AAAA}}{(l^2-m_A^2)^2} \right]\nonumber\\
&& +2\times\left[\frac{2V_{AA1}V_{AA12}}{[(l+p)^2-m_A^2](l^2-m_1^2)} + \frac{4V_{AA2}V_{AA22}}{[(l+p)^2-m_A^2](l^2-m_2^2)}\right] \nonumber\\
&& +\left[\frac{4 V_{112}V_{AA1}^2}{[(l+p)^2-m_A^2](l^2-m_1^2)^2} + \frac{2\times 4 V_{122}V_{AA1} V_{AA2}}{[(l+p)^2-m_A^2](l^2-m_1^2)(l^2-m_2^2)} \right. \nonumber\\
&&\left. +  \frac{12 V_{222}V_{AA2}^2}{[(l+p)^2-m_A^2](l^2-m_2^2)^2} \right]\nonumber\\
&& + \left[\frac{4V_{AA1}^2 V_{AA2}}{[(l+p)^2-m_A^2]^2(l^2-m_1^2)} + \frac{4V_{AA2}^3}{[(l+p)^2-m_A^2]^2 (l^2-m_2^2)}\right]\,,
\end{eqnarray}
Note that we have kept the momentum $p$ for external DM states while defining $\Delta V_{AA1}$ and $\Delta V_{AA2}$. 
The above 1PI irreducible diagrams are the basic ingredients for constructing more elaborated one-loop Feynman diagrams.

\begin{figure}[ht]
\centering
\includegraphics[width=\textwidth]{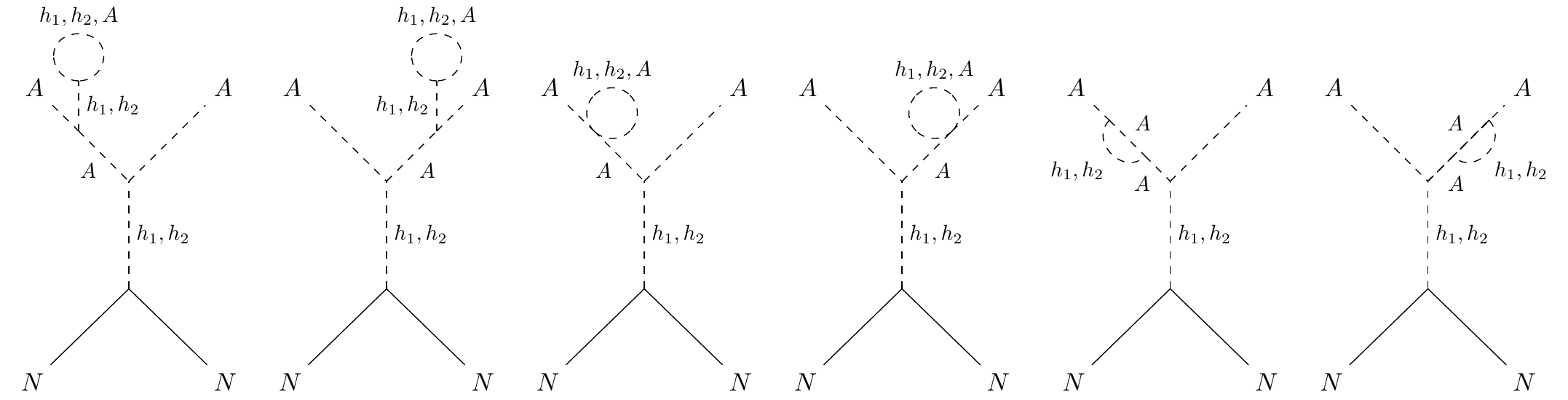}
\caption{One-loop diagrams with external $A$-line corrections}\label{FigLE}
\end{figure}

First of all, it is easy to write down the contributions to ${\cal F}$ from the one-loop external $A$ corrections shown in Fig.~\ref{FigLE}
\begin{eqnarray}
{\cal F}_e &=& \frac{2i}{p^2 - m_A^2}\left[ -i\Delta m_A^2 + \frac{2iV_{AA1}\Delta t_1}{m_1^2} + \frac{2iV_{AA2}\Delta t_2}{m_2^2} \right]{\cal F}_0 = 0\,,
\end{eqnarray}
where we have kept the same external $A$ momentum, $p$, which implies that the limit of zero momentum transfer was assumed.

\begin{figure}[ht]
\centering
\includegraphics[width=0.6\textwidth]{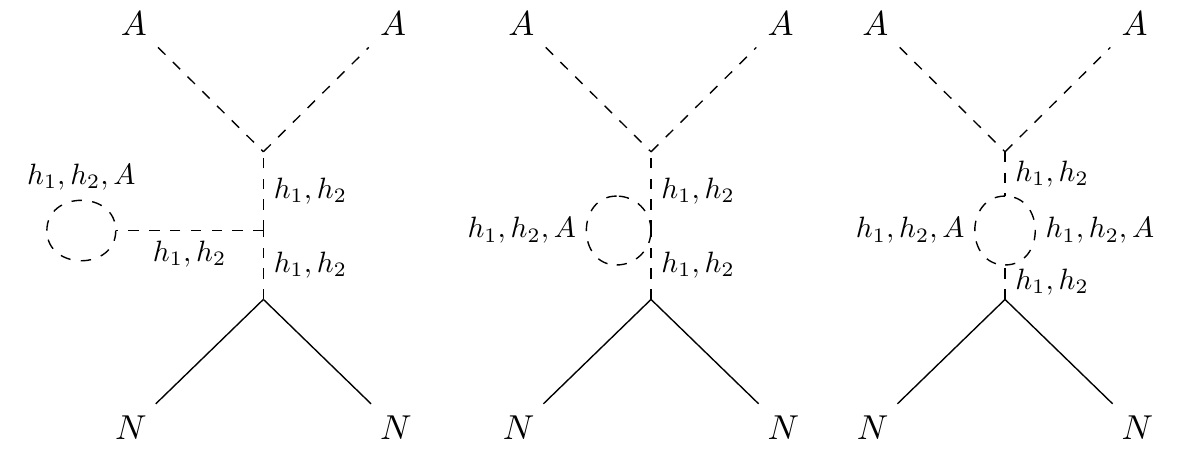}
\caption{One-loop diagrams with internal $h_{1,2}$ propagator corrections}\label{FigLI}
\end{figure}

\begin{figure}[ht]
\centering
\includegraphics[width=\textwidth]{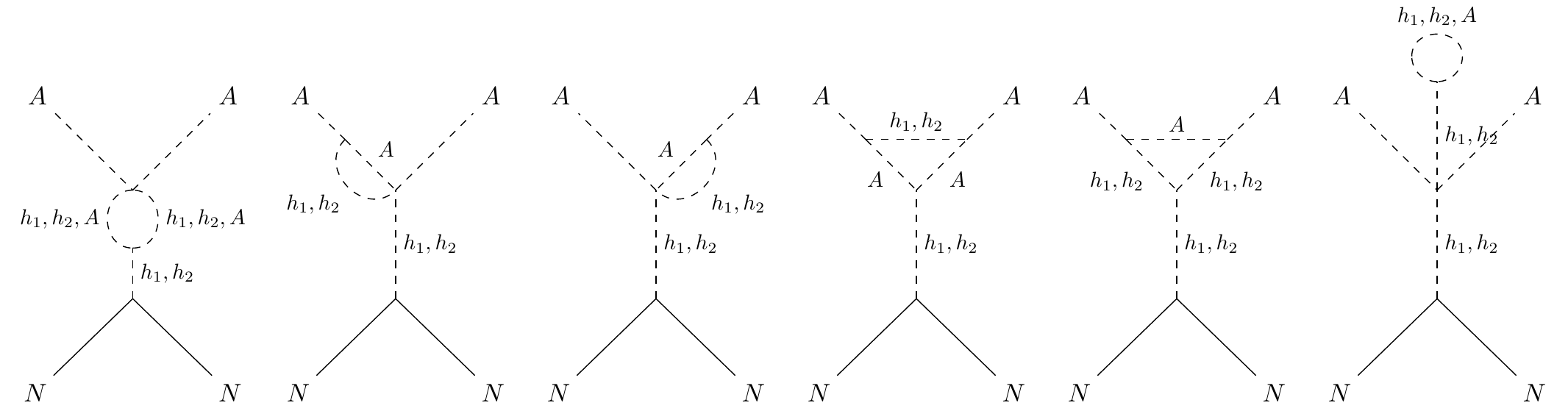}
\caption{One-loop diagrams with vertices $AAh_1$ and $AAh_2$ corrections}\label{FigLV}
\end{figure}

The remaining one-loop contributions are shown in Figs.~\ref{FigLI} and \ref{FigLV}. 
The reducible contributions to the vertices $AAh_1$ and $AAh_2$ due to the internal $h_{1,2}$ propagator corrections are given by
\begin{eqnarray}\label{Vi}
-i V_{AA1\,i}^{(1)} &=& iV_{AA1} \frac{\Delta m_1^2}{m_1^2} + iV_{AA2} \frac{\Delta m_{12}^2}{m_2^2} \nonumber\\
&&  -\frac{6iV_{AA1}V_{111} \Delta t_1}{m_1^4} -\frac{2 i V_{AA1} V_{112} \Delta t_2}{m_1^2 m_2^2}  -\frac{2 i V_{AA2} V_{112} \Delta t_1}{m_1^2 m_2^2} - \frac{2i V_{AA2} V_{122} \Delta t_2}{m_2^4}\,, \nonumber\\
-iV_{AA2\,i}^{(1)} &=& iV_{AA1} \frac{\Delta m_{12}^2}{m_1^2} + iV_{AA2} \frac{\Delta m_2^2}{m_2^2}\\
&&-\frac{2iV_{AA1} V_{112} \Delta t_1}{m_1^4} - \frac{2iV_{AA1}V_{122}\Delta t_2}{m_1^2 m_2^2} -\frac{2iV_{AA2}V_{122} \Delta t_1}{m_1^2 m_2^2} - \frac{6iV_{AA2}V_{222}\Delta t_2}{m_2^4}\,,\nonumber
\end{eqnarray}
while the ones from the vertex corrections are as follows:
\begin{eqnarray}
-i V_{AA1\,v}^{(1)} &=&  -i\Delta V_{AA1}  + \frac{2i V_{AA11}\Delta t_1}{m_1^2} + \frac{iV_{AA12}\Delta t_2}{m_2^2}\,,\nonumber\\
-i V_{AA2\,v}^{(1)} &=&  -i\Delta V_{AA2}  + \frac{i V_{AA12}\Delta t_1}{m_1^2} + \frac{2iV_{AA22} \Delta t_2}{m_2^2}\,.
\end{eqnarray}

Thus, the total one-loop contributions to the factor ${\cal F}$ is given by
\begin{eqnarray}\label{FO}
{\cal F} &=& \frac{ (V_{AA1\,i}^{(1)}+V_{AA1\,v}^{(1)})c_\alpha}{m_1^2} - \frac{ (V_{AA2\,i}^{(1)}+V_{AA2\,v}^{(1)})s_\alpha }{m_2^2}\nonumber\\
&=& \frac{ i s_{2\alpha} (m_1^2 -m_2^2) }{8 v_H v_S^3 m_1^2 m_2^2 }\int\frac{d^4l}{(2\pi)^4} \left[ \frac{{\cal A}_1 (l\cdot p) }{(l^2-m_1^2)(l^2-m_2^2)[(l+p)^2-m_A^2]} \right. \\
&& + \left. \frac{{\cal A}_2 (l\cdot p) }{(l^2-m_1^2)^2(l^2-m_2^2)[(l+p)^2-m_A^2]} + \frac{{\cal A}_3 (l\cdot p) }{(l^2-m_1^2)(l^2-m_2^2)^2[(l+p)^2-m_A^2]} \right]\nonumber
\end{eqnarray}
where the coefficients ${\cal A}_i$ are defined as follows
\begin{eqnarray}
{\cal A}_1 &\equiv & 2 (m_1^2 s_\alpha^2 +m_2^2 c_\alpha^2) ( 2 m_1^2 v_H s_\alpha^2 + 2 m_2^2 v_H c_\alpha^2 - m_1^2 v_S s_{2\alpha} + m_2^2 v_S s_{2\alpha})\,,\nonumber\\
{\cal A}_2 &\equiv & -2m_1^4 s_\alpha [(m_1^2 + 5m_2^2)v_S c_\alpha - (m_1^2 -m_2^2) (v_S c_{3\alpha}+4v_H s_\alpha^3) ]\,,\\
{\cal A}_3 &\equiv & 2 m_2^4 c_\alpha [(5m_1^2+m_2^2)v_S s_\alpha - (m_1^2 -m_2^2)(v_S s_{3\alpha} + 4 v_H c_\alpha^3)]\,.
\nonumber
\end{eqnarray}
Note that in the derivation of Eq.~(\ref{FO}) we have used the tree-level relations from Eq.~(\ref{ConsistCond}) and the DM particle on-shell condition $p^2 = m_A^2$.

We can utilize the Passarino-Veltman $C$ and $D$ functions as defined in Refs.~\cite{Passarino:1978jh,Denner:1991kt,Hahn:1998yk} to further reduce the expression of ${\cal F}$ to be
\begin{eqnarray}\label{F1L}
{\cal F} &=& -\frac{s_{2\alpha}(m_1^2-m_2^2)}{128\pi^2 v_H v_S^3 m_1^2 m_2^2} p^\mu [{\cal A}_1  C_\mu(0,p^2, p^2, m_1^2, m_2^2, m_A^2)\nonumber\\
&& + {\cal A}_2  D_\mu (0,0,p^2, p^2, 0, m_A^2, m_1^2, m_1^2, m_2^2,m_A^2) \nonumber\\
&&  + {\cal A}_3 D_\mu(0,0,p^2, p^2, 0, m_A^2, m_1^2, m_2^2, m_2^2,m_A^2) ]\nonumber\\
&=& -\frac{s_{2\alpha}(m_1^2-m_2^2) m_A^2}{128\pi^2 v_H v_S^3 m_1^2 m_2^2}  [{\cal A}_1  C_2(0,m_A^2, m_A^2, m_1^2, m_2^2, m_A^2)\nonumber\\
&& + {\cal A}_2  D_3 (0,0,m_A^2, m_A^2, 0, m_A^2, m_1^2, m_1^2, m_2^2,m_A^2) \nonumber\\
&&  + {\cal A}_3 D_3(0,0,m_A^2, m_A^2, 0, m_A^2, m_1^2, m_2^2, m_2^2,m_A^2) ]\,,
\end{eqnarray}
where we have used $p^2 = m_A^2$ and the following identity
\begin{eqnarray}
C_\mu(0,p^2, p^2, m_1^2, m_2^2, m_A^2) = p_\mu C_2(0,p^2, p^2, m_1^2, m_2^2, m_A^2)\,,
\end{eqnarray}
as well as the similar identities for $D$ functions. As anticipated earlier, this expression shows that the one-loop DM-nucleon scattering amplitude is finite in the zero momentum-transfer limit. Moreover, since ${\cal F}$ is proportional to $m_A^2$ and the $C_2$ and $D_3$ functions behave at most as $\sim \ln m_A$ in the limit $m_A\to 0$, the amplitude vanishes (as expected) in the limit $m_A \to 0$. It is highly non-trivial to satisfy both conditions at the same time, therefore this is an important test of our results. 

%%%%%%%%%%%%%%%%%%%%%%%%%%%%%%%%%%%%%%%%%%%%%%%%%%%%%%%%%%%%%%%%%%%%%%%%%%%%%%%%%%%%%%%%%%%%
\section{Numerical Studies}\label{Sec_Numerical}
\begin{figure}[ht]
\centering
\includegraphics[scale = 0.7]{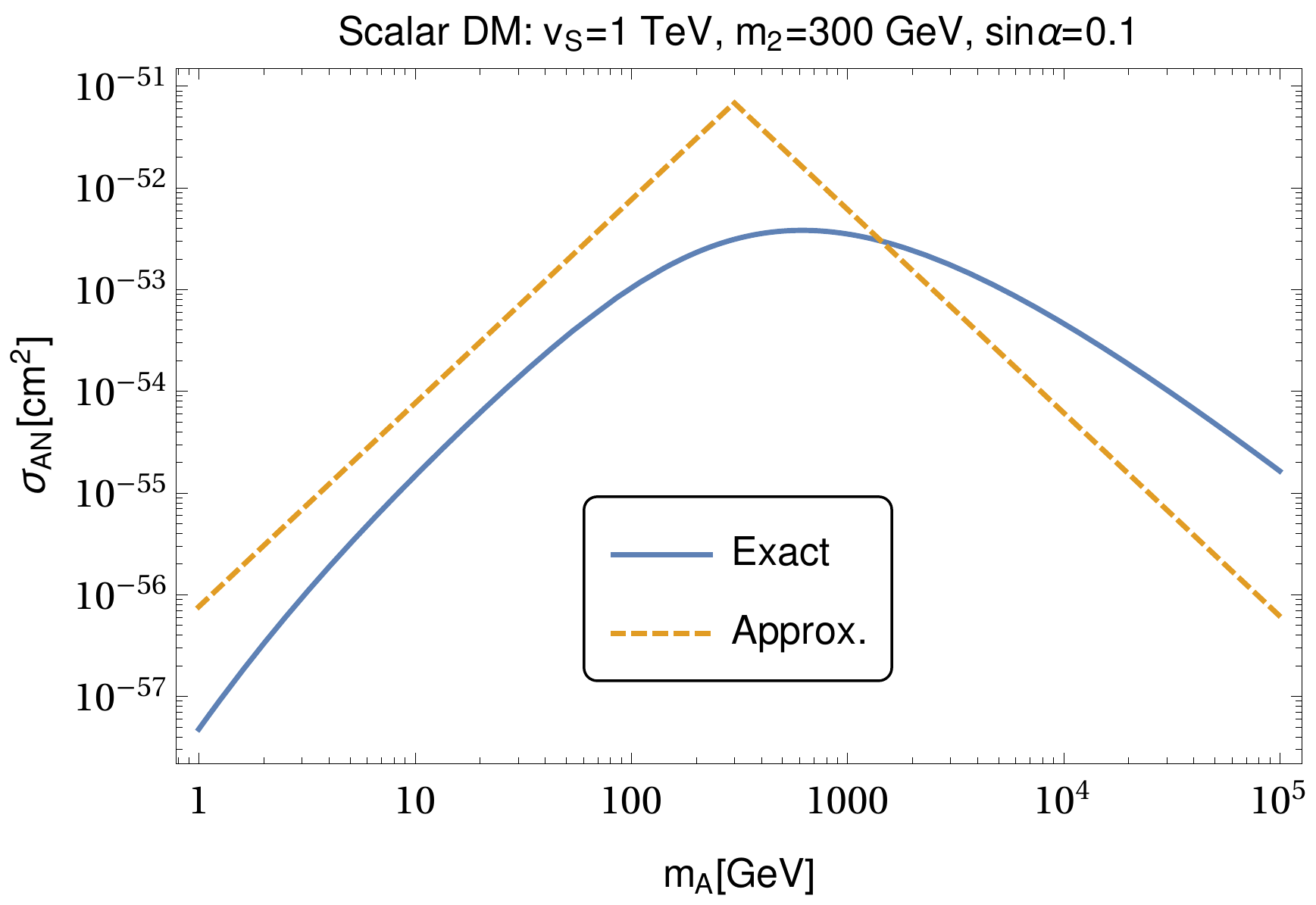}
\caption{The DM-nucleon scattering cross section $\sigma_{AN}$ as the function of the DM mass $m_A$. The blue solid curve represents the exact leading-order one-loop contribution in the limit of vanishing DM velocity, while the yellow dashed curve displays the approximate results proposed in Ref.~\cite{Gross:2017dan}. }\label{FigXS}
\end{figure}

Having the explicit expression of the one-loop DM-nucleon recoiling cross section $\sigma_{AN}^{(1)}$ in Eq.~(\ref{XS1L}) with its loop function ${\cal F}$ in Eq.~(\ref{F1L}), we can calculate the magnitude of the DM-nucleon cross section with typical model parameters. In this section, we take $v_S = 1~{\rm TeV}$, $m_2 = 300$~GeV, $s_\alpha = 0.1$, while leaving the DM mass varying freely. Note that we have reduced the final analytic expression for ${\cal F}$ in terms of the Passarino-Veltman functions, so that it is easy to calculate it numerically  adopting the package {\tt LoopTools}~\cite{Hahn:1998yk}. The final result is displayed in Fig.~\ref{FigXS} as the smooth solid blue curve. We note that, for the given set of parameters, the DM-nucleon scattering cross section varies between $10^{-58}~{\rm cm}^2$ and $10^{-52}~{\rm cm}^2$ when the DM mass $m_A$ is in the range of $1 \;-\; 10^5$~GeV. For the same set of the parameters the curve has a maximum value of  $\sigma^{(1)}_{AN\, {\rm max}} \sim 3\times 10^{-53}~{\rm cm}^2$ for $m_A\sim 630$~GeV. This should be compared with the tree-level contribution at the leading order of the DM velocity given in Eq.~(\ref{XStree}), which predicts $\sigma_{AN}^{{\rm tree}} \sim 10^{-69}$ -- $10^{-65}~{\rm cm}^2$ with the same set of parameters. Thus, we can conclude that the leading-order DM-nucleon cross section is provided by the one-loop contributions at vanishing DM velocity, rather than the finite velocity corrections.

In contrast, we also show as the dashed yellow curve in Fig.~\ref{FigXS} the following approximation proposed in Ref.~\cite{Gross:2017dan} as an estimate of the one-loop cross section
\begin{eqnarray}\label{approx}
\sigma^{(1)}_{AN} \approx \left\{\begin{array}{cc}
\frac{s_\alpha^2}{64\pi^5} \frac{m_N^4 f_N^2}{m_1^4 v_H^2} \frac{m_2^8}{m_A^2 v_S^6}\,, & m_A \geq m_2 \\
\frac{s_\alpha^2}{64\pi^5} \frac{m_N^4 f_N^2}{m_1^4 v_H^2} \frac{m_2^4 m_A^2}{v_S^6}\,, & m_A \leq m_2 \\
\end{array}\right.\,.
\end{eqnarray} 
It is clear that when $m_A$ lies below 1~TeV, the approximation is about one-order larger than the exact result, while, if $m_A \gg 1$~TeV, the exact $\sigma^{(1)}_{AN}$ is almost one-order higher. Nevertheless, these two curves share almost the same scaling behaviour in the limits of very small and very large DM masses, which are reflected by the same slopes in the plot. Furthermore, both are well below the currently most stringent experimental limit of ${\cal O}(10^{-47})~{\rm cm}^2$ given by the XENON1T Collaboration. Therefore, the conclusion given in Refs.~\cite{Gross:2017dan,Azevedo:2018oxv} that the DM direct detections does not impose any relevant constraints on the present model does not change. In particular, the available parameter space given in Ref.~\cite{Gross:2017dan,Azevedo:2018oxv} is still the same, and would not change by using the exact formulae 
presented in Eqs.~(\ref{XS1L}) and (\ref{F1L}) in the parameter scan.

\begin{figure}[ht]
\centering
\includegraphics[scale = 0.7]{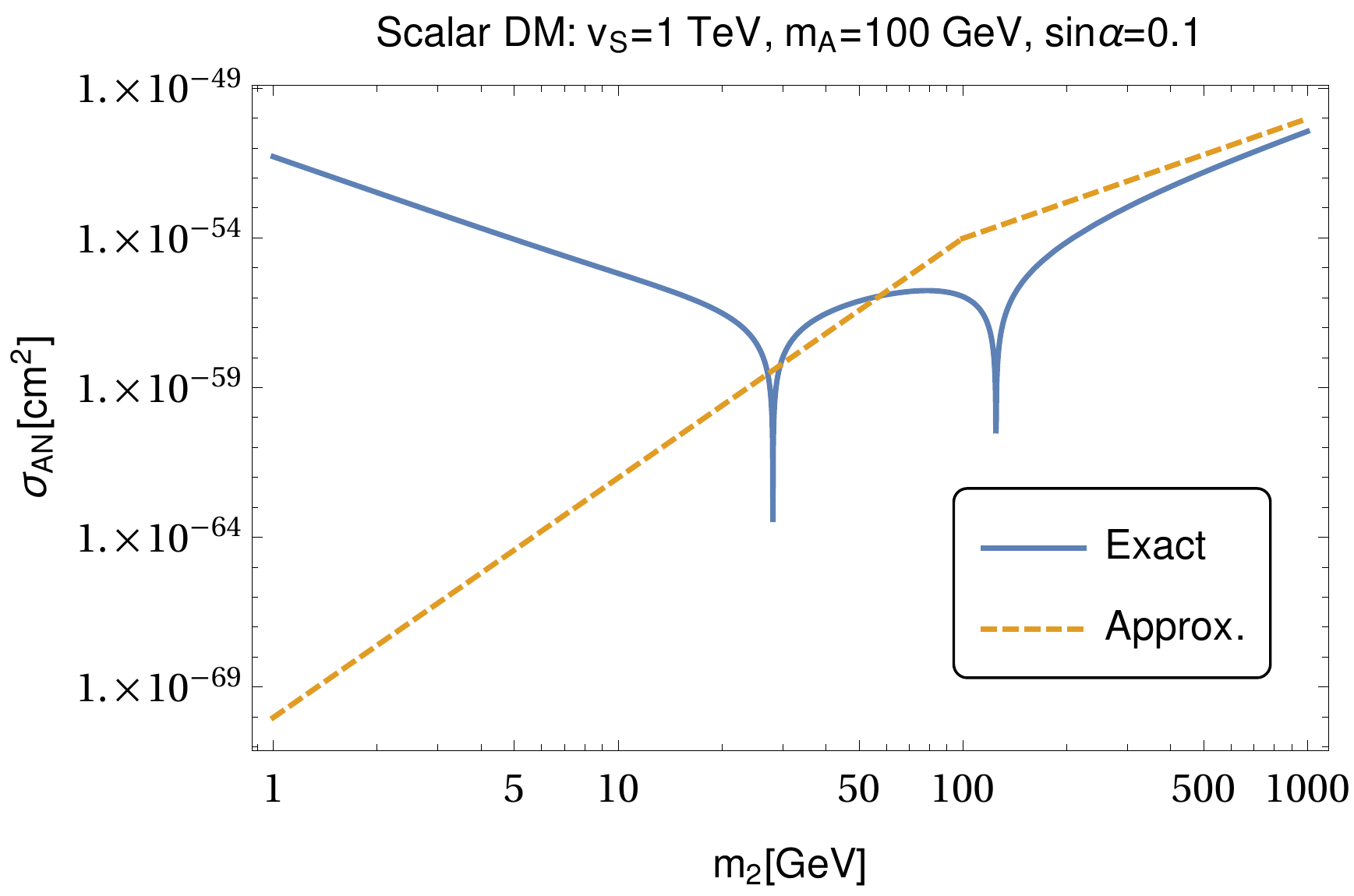}
\caption{The DM-nucleon scattering cross section $\sigma_{AN}$ as the function of $m_2$. The blue solid curve represents the exact leading-order one-loop contribution in the limit of vanishing DM velocity, while the yellow dashed curve displays the approximate results proposed in Ref.~\cite{Gross:2017dan}. }\label{FigXS1}
\end{figure}

In Fig.~\ref{FigXS1}, we also show the DM-nucleon cross section as a function of the mass of non-SM-like Higgs boson $h_2$ with a fixed DM mass. We can see that the approximation, shown as the yellow dashed curve, substantially deviates from the exact formula in Eqs.~(\ref{XS1L}) and (\ref{F1L}) drawn as the blue solid curve. The one-loop result shows a much richer structure as the $h_2$ mass increases, rather than the simple scaling law predicted in Eq.~(\ref{approx}). In particular, two dips appear in the exact calculation. It is easy to see that one of them is located exactly at the point where $m_2=m_1$ corresponding to the vanishing of the factor $(m_1^2-m_2^2)$ in Eq.~(\ref{F1L}). Another dip appearing at around $m_2 \sim 30$~GeV is caused by accidental cancellation between loop integrals. The location of this dip varies with the set of parameters chosen and is a combination
of all input parameters, the mass of the scalars, the angle $\alpha$ and $v_S$ . Furthermore, note that when the $h_2$ mass is very small, the DM-nucleon cross section decreases as $m_2$ grows, in contrast with what is predicted by the approximate expression. On the other hand, when $m_2$ becomes much larger than the DM mass $m_A = 100$~GeV, the two curves approach each other, indicating that the approximation becomes valid only in this region. Finally we note that there is no difference between the approximate and the exact expression in the behaviour with the angle $\alpha$ and with the VEV $v_S$.

%%%%%%%%%%%%%%%%%%%%%%%%%%%%%%%%%%%%%%%%%%%%%%%%%%%%%%%%%%%%%%%%%%%%%%%%%%%%%%%%%%%%%%%%%%%%%%%
\section{Conclusion}\label{Sec_Conc}

In this work we have computed the one-loop electroweak contribution to DM-nucleon scattering, at zero momentum transfer, in a complex singlet extension of the SM with a softly
broken $U(1)$ symmetry. It has been shown in Refs.~\cite{Gross:2017dan,Azevedo:2018oxv} that in such a simple extension of the SM 
with an extra complex scalar $S$ and with a softly broken $U(1)$ symmetry, the pseudo-Goldstone component $A$ becomes the DM candidate
 and the tree-level contributions to the DM-nucleon recoiling cross section vanishes in the limit of zero momentum transfer. 
Therefore, the model has the attractive feature that the DM-nucleon cross section is naturally suppressed. Hence it is important to verify how 
large are one-loop contributions to the DM-nucleon scattering in this model. The calculation of these corrections,
 in the limit of zero-momentum transfer, was the main goal of this work.
%In the light of the importance of this model which can naturally avoid the strong constraints from the DM direct detections, we have carefully calculated the one-loop contribution to the DM-nucleon cross section in the limit of the vanishing momentum transfer. 
We have shown that, for typical parameter choices, this one-loop contribution is 10 orders of magnitude larger than the finite-velocity or finite-momentum-transfer corrections at tree-level. Therefore, we explicitly prove the expectation that the leading-order $\sigma_{AN}$ indeed arises at the one-loop level. Furthermore, with the explicit analytic expression of $\sigma_{AN}$ given in Eqs.~(\ref{XS1L}) and (\ref{F1L}), we show
that the one-loop contribution is finite and approaches zero in the limit of vanishing DM mass. Finally, it has been shown that the DM-nucleon cross section is typically well below ${\cal O}(10^{-50}~{\rm cm}^2)$, which is much lower than the most stringent experimental upper bounds of ${\cal O}(10^{-47})$ from XENON1T. This indicates that this model suppresses the DM direct detection signals so effectively that it is not constrained at all by this kind of experiments. Still, these radiative corrections will be important for the next generation of DM direct detection experiments, when the values of the cross sections that can be probed will reach the level of the one-loop
result presented in this work. Finally we found accidental blind spots at the one-loop level, that is, points for which the DM-nucleon scattering cross section is still vanishingly small. These blind spots appear for 
given combination of parameters for which a next order calculation would be needed.

%%%%%%%%%%%%%%%%%%%%%%%%%%%%%%%%%%%%%%%%%%%%%%%%%%%%%%%%%%%%%%%%%%%%%%%%%%%%%%%%%%%%%%%%%%%%%
\appendix
\section{Tree-level interacting vertices}\label{appA}
By expanding the tree-level potential in Eq.~(\ref{potential}) in terms of the physical mass eigenstates $h_{1,2}$ and $A$, the tree-level triple- and quartic-scalar vertices can be written as follows,
\begin{eqnarray}\label{vertices}
{\cal V}_{\rm int} &=& V_{111} h_1^3 + V_{112} h_1^2 h_2 + V_{122} h_1 h_2^2 + V_{222} h_2^3 + V_{AA1} h_1 A^2 + V_{AA2} h_2 A^2 \nonumber\\
&& + V_{1111} h_1^4 + V_{1112} h_1^3 h_2 + V_{1122} h_1^2 h_2^2 + V_{1222} h_1 h_2^3 + V_{2222} h_2^4\\
&& + V_{AA11} h_1^2 A^2 + V_{AA12} h_1 h_2 A^2 + V_{AA22} h_2^2 A^2 + V_{AAAA} A^4 \,. \nonumber
\end{eqnarray}
Note that only even powers of $A$ appear in the above interaction vertices which manifests the DM nature of $A$. The coefficients of the above vertices are listed below for reference,
\begin{eqnarray}
V_{111} &=& c_\alpha^3 \lambda_H v_H  + \frac{1}{2} s_\alpha c_\alpha^2 \kappa v_S  + \frac{1}{2} s_\alpha^2 c_\alpha \kappa v_H  + s_\alpha^3 \lambda_S v_S \,,\nonumber\\
V_{112} &=& \frac{1}{2}c_\alpha^3 \kappa v_S + s_\alpha c_\alpha^2 \kappa v_H -3 s_\alpha c_\alpha^2 \lambda_H v_H - s_\alpha^2 c_\alpha \kappa v_S + 3 s_\alpha^2 c_\alpha \lambda_S v_S - \frac{1}{2} s_\alpha^3 \kappa v_H \,,\nonumber\\
V_{122} &=& \frac{1}{2} c_\alpha^3 \kappa v_H - s_\alpha c_\alpha^2 \kappa v_S +3 s_\alpha c_\alpha^2 \lambda_S v_S -s_\alpha^2 c_\alpha \kappa v_H + 3 s_\alpha^2 c_\alpha \lambda_H v_H + \frac{1}{2} s_\alpha^3 \kappa v_S\,,\\
V_{222} &=& c_\alpha^3 \lambda_S v_S - \frac{1}{2} s_\alpha c_\alpha^2 \kappa v_H + \frac{1}{2} s_\alpha^2 c_\alpha \kappa v_S - s_\alpha^3 \lambda_H v_H\,,\nonumber\\
V_{AA1} &=& s_\alpha \lambda_S v_S + \frac{1}{2} c_\alpha \kappa v_H \,,\nonumber\\
V_{AA2} &=& c_\alpha \lambda_S v_S - \frac{1}{2} s_\alpha \kappa v_H \,,\nonumber
\end{eqnarray}
\begin{eqnarray}
V_{1111} &=& \frac{1}{4} (c_\alpha^4 \lambda_H + s_\alpha^2 c_\alpha^2 \kappa + s_\alpha^4 \lambda_S)\,,\nonumber\\
V_{1112} &=& \frac{1}{2}s_\alpha c_\alpha^3 \kappa - s_\alpha c_\alpha^3 \lambda_H -\frac{1}{2} s_\alpha^3 c_\alpha \kappa + s_\alpha^3 c_\alpha \lambda_S\,,\nonumber\\
V_{1122} &=& \frac{1}{4} (c_\alpha^4 \kappa + s_\alpha^4 \kappa - 4\kappa s_\alpha^2 c_\alpha^2 \kappa + 6 s_\alpha^2 c_\alpha^2 \lambda_H + 6 s_\alpha^2 c_\alpha^2 \lambda_S)\,,\nonumber\\
V_{1222} &=& -\frac{1}{2}s_\alpha c_\alpha^3 \kappa + s_\alpha c_\alpha^3 \lambda_S + \frac{1}{2} s_\alpha^3 c_\alpha \kappa - s_\alpha^3 c_\alpha \lambda_H\,,\nonumber\\
V_{2222} &=& \frac{1}{4} (c_\alpha^4 \lambda_S + s_\alpha^2 c_\alpha^2 \kappa + s_\alpha^4 \lambda_H)\,,\\
V_{AA11} &=& \frac{1}{4} (2 s_\alpha^2 \lambda_S + c_\alpha^2 \kappa )\,,\nonumber\\
V_{AA12} &=& -\frac{1}{2} s_\alpha c_\alpha \kappa + s_\alpha c_\alpha \lambda_S \,,\nonumber\\
V_{AA22} &=& \frac{1}{4}  (2 c_\alpha^2 \lambda_S + s_\alpha^2 \kappa )\,,\nonumber\\
V_{AAAA} &=& \frac{\lambda_S}{4}\,.\nonumber
\end{eqnarray}

\section{Analytic Expressions of $C$ and $D$ Functions}
In this appendix, we try to give the explicit expressions for the $C$ and $D$ functions given in Eq.~(\ref{F1L}). 

\begin{eqnarray}\label{Cmu}
C_\mu &=& \int \frac{d^4 l}{(2\pi)^4} \frac{l_\mu}{(l^2-m_1^2)(l^2-m_2^2)[(l+p)^2-m_A^2]}\nonumber\\
&=& \frac{1}{m_1^2 - m_2^2}\left({\cal I}^{1}_\mu(m_1^2, m_A^2) - {\cal I}^{1}_\mu(m_2^2, m_A^2)\right) \,,
\end{eqnarray}
where the function ${\cal I}^1_\mu$ is defined as follows
\begin{eqnarray}\label{I1}
&&{\cal I}^1_{\mu} (m_1^2, m_A^2) \equiv  \int\frac{d^4 l}{(2\pi)^4} \frac{l_\mu}{(l^2-m_1^2)[(l+p)^2-m_A^2]} \nonumber\\
&=& -\frac{i p_\mu}{16\pi^2}\frac{1}{2} \left\{ \left(\frac{2}{\epsilon}-\gamma +{\rm ln}\frac{\mu^2}{m_A^2}\right) + \left[ 1+ (x_1 + x_2) + x_1^2 \ln\frac{x_1 - 1}{x_1} + x_2^2 \ln\frac{x_2 -1}{x_2} \right] \right\} \,,\nonumber\\
\end{eqnarray}
where we have defined the symbols $x_{1,2} \equiv  (m_1^2 \pm \sqrt{m_1^4 - 4m_1^2 m_A^2})/(2m_A^2)$ to denote the two roots of the equation $x^2 - m_1^2 x /m_A^2 +m_1^2/m_A^2 = 0$. We also have used the dimensional regularization to regularize the UV divergence in ${\cal I}^1_\mu$. Note that Eq.~(\ref{I1}) is only valid when $m_1^2 > 4 m_A^2$. With the explicit expression in Eqs.~(\ref{Cmu}) and (\ref{I1}), we can easily prove the finiteness of $C_\mu$ in the $m_A^2 \to 0$ limit. In fact, in the limit of $m_A^2/m_1^2 \to 0$, $I^{1}_{\mu}(m_1^2, m_A^2)$ can be reduced to
\begin{eqnarray}
I^1_{\mu} (m_1^2, m_A^2) \to -\frac{ip_\mu}{32\pi^2} \left(\frac{2}{\epsilon} -\gamma + \ln \frac{\mu^2}{m_1^2} + \frac{3}{2}\right).
\end{eqnarray}
Therefore, the $C$ function defined in the main text is given as follows in the $m_A^2 \to 0$ limit
\begin{eqnarray}\label{Climit}
C_\mu = \frac{ip_\mu}{32\pi^2}\frac{\ln (m_2^2/m_1^2)}{m_2^2 -m_1^2}\,,
\end{eqnarray}
which is obviously finite. 

We can address the D functions in Eq.~(\ref{F1L}) in the similar way 
\begin{eqnarray}\label{D1}
&& D_\mu (0,0,p^2, p^2, 0, m_A^2, m_1^2, m_1^2, m_2^2,m_A^2)  \nonumber\\
&\equiv & \int \frac{d^4 l}{(2\pi)^4} \frac{l_\mu}{(l^2-m_1^2)^2(l^2-m_2^2)[(l+p)^2-m_A^2]}\nonumber\\
&=& \frac{1}{m_1^2 - m_2^2}\left[{\cal I}^{2}_\mu(m_1^2, m_A^2) - { C}_\mu(0, p^2, p^2,m_1^2, m_2^2, m_A^2)\right] \,,
\end{eqnarray}
where 
\begin{eqnarray}
&&{\cal I}^2_\mu (m_1^2, m_A^2) \equiv \int \frac{d^4 l}{(2\pi)^4} \frac{l_\mu}{(l^2 -m_1^2)^2 [(l+p)^2-m_A^2]}\nonumber\\
&=& -\frac{ip_\mu}{16\pi^2 m_A^2} \left[1+ \frac{x_1 (x_1 -1)}{x_1-x_2}\ln\frac{x_1 -1}{x_1} - \frac{x_2 (x_2 -1)}{x_1-x_2}\ln\frac{x_2 -1}{x_2} \right]\,,
\end{eqnarray}
in which $x_{1,2}$ is defined as before. By taking the zero DM mass limit, ${\cal I}^2_\mu(m_1^2,m_A^2)$ can be reduced to
\begin{eqnarray}
{\cal I}^2_\mu(m_1^2, m_A^2) \to \frac{ip_\mu}{32\pi^2 m_1^2}\,,
\end{eqnarray}
which can also obtained by taking the limit of $m_2^2 \to m_1^2$ in Eq.~(\ref{Climit}). Since $C$ function is finite in the same limit as proved earlier, the $D$ function is also finite in this limit. The same result can be applied to another $D$ function since it is obtained from Eq.~(\ref{D1}) via the exchange of $m_1 \leftrightarrow m_2$.

%%%%%%%%%%%%%%%%%%%%%%%%%%%%%%%%%%%%%%%%%%%%%%%%%%%%%%%%%%%%%%%%%%%%%%%%%%%%%%%%%%%%%%%%%%%%%%%
\section*{Acknowledgments}
%This work is supported by the National Science Centre (Poland) research project, decision DEC-2014/15/B/ST2/00108.
This work is supported in part by the National Science Centre (Poland), research
projects no 2014/15/B/ST2/00108, no 2017/25/B/ST2/00191 and a HARMONIA project
under contract UMO-2015/18/M/ST2/00518 (2016-2019).
%\section*{References}

\end{document}